 \documentstyle[12pt]{article}
\def\ll{\label}
\def\re{\ref}
\def\c{\cite}

\def\r1{(\ref{$1})}

\def\ti{\tilde}

\def\ba{\begin{array}{c}}

\def\ea{\end{array}}

\def\ni{\noindent}
\def\si{\sigma}

\def\De{\Delta}

\def\ov{\over}
\def\ha{{1\over 2}}

\def\l{\left}
\def\l({\left(}
\def\r){\right)}
\def\r{\right}

\def\la{\lambda}
\def\al{\alpha}

\def\be{\begin{equation}}
\def\bc{\begin{center}}
\def\ec{\end{center}}
\def\bit{\begin{itemize}}
\def\eit{\end{itemize}}
\def\ee{\end{equation}}
\def\ed{\end{document}}
\def\bea{\begin{eqnarray}}
\def\eea{\end{eqnarray}}
\def\efr{\end{flushright}}

\begin{document}
\title{Construction of variable mass sine-Gordon  and other novel 
 inhomogeneous  quantum
 integrable  models
}

\author{
Anjan Kundu \footnote {email: anjan@tnp.saha.ernet.in} \\  
  Saha Institute of Nuclear Physics,  
 Theory Group \\
 1/AF Bidhan Nagar, Calcutta 700 064, India.
 }
\maketitle
\vskip 1 cm

\begin{abstract} 
The inhomogeneity of the media or the external forces usually destroy
the integrability of a system. We propose a 
systematic construction of  a class of quantum models, 
which retains their exact integrability inspite of their 
explicit  inhomogeneity.
Such models include
 variable mass sine-Gordon model, 
cylindrical  NLS,  spin chains with impurity, 
inhomogeneous Toda chain, the  Ablowitz-Ladik model etc.
\medskip
\end{abstract}

\smallskip

  
\section{Introduction}
\setcounter{equation}{0}

The  physical systems often encounter  the inhomogeneity in the form of
impurities, defects or the density fluctuations in the media or it
 can enter as variable magnetic fields or other external
forces.
 Such  inhomogeneities 
can depend on space as well as time variables and can appear as 
explicit space-time dependent coefficients in the Hamiltonian. 
They
 usually destroy the integrability of a system
 making        its analytic study almost impossible \c{msg}.
However there are several examples  of 
classical  models like  deformed MKDV \c{burtsev} or NLS \c{CL},
spherical or cylindrical symmetric 
NLS  \c{rl}, inhomogeneous Ablowitz-Ladik (AL)
model \c{konotop} etc.
 where the exact integrability could be retained along 
with  their Lax operators, inspite of the presence of  space-time
dependent coefficients in their evolution equations.
Nevertheless, in case of  quantum models there seems to be no
 systematic 
attempts to explore such possibilities,
 except certain construction for  the impurity chains \c{impchain} .
%

We propose here  a  scheme for introducing inhomogeneities in the known
quantum integrable models, which retains their integrability and allows
explicit construction of their $R$-matrix as well as
 the Lax operators. This systematic  scheme is based on a novel quadratic
 algebra derived from the quantum Yang--Baxter
equation,
 where its Casimir operators play the role of the
inhomogeneity parameters and the proper realisation of other
generators  construct the Lax operator of the model. The $R$-matrix
simply corresponds to that of the standard homogeneous model. Applying this
procedure one first constructs quantum models with inhomogeneity on discrete
lattices, which preserves the exact integrability
 and then taking the  continuum limit  builds the corresponding  field
models. Thus one obtains a  series of new quantum integrable 
inhomogeneous models    like
a  variable mass quantum sine-Gordon model, cylindrical quantum
  NLS,  
inhomogeneous Toda  and   Ablowitz Ladik chains. It also provides a
different way of introducing integrable impurities in 
 the spin chain models. 

\section{The generating scheme}
\setcounter{equation}{0}

    We start with the quadratic  algebra \c{kunprl99} 
\be
 [S^3,S^{\pm}] = \pm S^{\pm} , \ \ \ [ S^ {+}, S^{-} ] =
 \left ( M^+\sin (2 \al S^3) + {M^- } \cos
( 2 \al S^3  ) \right){1 \over \sin \al}, \quad  [M^\pm, \cdot]=0,
\ll{nlslq2a}\ee
and the quantum L-operator
\be
L_t{(\xi)} = \left( \begin{array}{c}
  \xi{c_1^+} e^{i \al S^3}+ \xi^{-1}{c_1^-}  e^{-i \al S^3}\qquad \ \ 
2 \sin \al  S^-   \\
    \quad  
2 \sin \al  S^+    \qquad \ \  \xi{c_2^+}e^{-i \al S^3}+ 
\xi^{-1}{c_2^-}e^{i \al S^3}
          \end{array}   \right), \quad
          \xi=e^{i \alpha \la}. \ll{nlslq2} \ee
where   
$ M^\pm=\pm   \sqrt {\pm 1} ( c^+_1c^-_2 \pm
c^-_1c^+_2 ) $ are  the Casimir operators 
 of algebra (\re {nlslq2a}).  
Taking   
 the well known  trigonometric   $R(\la)$-matrix solution \c{kulskly} 
along with (\re{nlslq2}) the quadratic algebra  (\re{nlslq2a}) 
 can be shown to be  equivalent to the 
 quantum Yang Baxter relation \c{kulskly} $R(\la-\mu) L(\la)\otimes  
L(\mu)= (I \otimes L(\mu) ) \otimes ( L(\la)\otimes I) R(\la-\mu)$.
Therefore the associated  L-operator 
(\re {nlslq2}) may serve as the  generating  Lax operator for  
the quantum integrable  models belonging to the relativistic or the
anisotropic class of models, the parameter $q=e^{i \al}$ 
playing the role of the deformation parameter \c{tarunpr99}. The    
integrable inhomogeneities are
 introduced in fact through different
  representations of the Casimir operators 
$c^\pm_a$ by choosing their  eigenvalues 
 as position and time  dependent functions.

At the undeformed limit $q=e^\al \to 1$ or equivalently at $\al \to 0$
all           the entries in the above scheme, i.e.
  algebra (\re {nlslq2a}),  L-operator 
(\re {nlslq2}) and the trigonometric $R$-matrix 
   are  reduced to   their corresponding rational forms.
The reduced algebra is simplified but still represents a quadratic algebra: 
\be  [ s^+ , s^- ]
=  2m^+ s^3 +m^-,\ \ \ \ 
  ~ [s^3, s^\pm]  = \pm s^\pm 
  \ll{k-alg} \ee
with $m^+=c_1^0c_2^0,\ \  m^-= c_1^1c_2^0+c_1^0c_2^1$ as
the new central elements.
Note that both  (\re {nlslq2a}) and (\re {k-alg})
  are Hopf algebras  with explicit coproduct structure, counit, antipode etc.
\c{tarunpr99}.

Due to $\al \to 0$ and $\xi \to 1+ \al 
\la $ the  L-operator   takes the form
 \be
L_r{(\la)} = \left( \begin{array}{c}
 {c_1^0} (\la + {s^3})+ {c_1^1} \ \ \quad 
  s^-   \\
    \quad  
s^+    \quad \ \ 
c_2^0 (\la - {s^3})- {c_2^1}
          \end{array}   \right), \ll{LK} \ee
 with   spectral parameter $\la$ and 
   the quantum
 $R$-matrix  is reduced  to its well known rational form \c{kulskly}. 
 Remarkably, our  scheme with these reduced entries  
belonging to the 
rational   class becomes  suitable  
for generating  quantum integrable nonrelativistic  
  models with inhomogeneity. 

\section{Inhomogeneous quantum integrable models}

\subsection{Variable mass sine-Gordon model}

Since this is a relativistic model we have to use the objects belonging to
the trigonometric class.
Through canonical operators $u,p$  a representation of 
(\ref{nlslq2a})
 may be given by 
\be
 S^3=u, \ \ \   S^+=  e^{-i p}g(u),\ \ \ 
 S^-=  g(u)e^{i p},
\ll{ilsg}\ee
where
the operator function
\be g (u)= \left ( 1-\sin \al u (M^+ \sin \al (u+1) )
 \right )^{\ha}  { 1 \ov \sin \al } \ll{g}\ee
By choosing  the  eigenvalues of the Casimirs as
$ M_j^+=- (\De m_j)^2, M_j^-=0$
and inserting (\re {ilsg}) in (\ref{nlslq2}) 
one gets a quantum integrable lattice model involving bosonic operators 
and the  inhomogeneity parameter $m_j. $ Comparing with the well known
result \c{korepinsg} we may conclude that the model thus constructed is 
a generalisation of the exact lattice version of the  quantum sine-Gordon
model.  
For going to the 
 continuum  limit we may scale $p$ by  lattice constant $\Delta$
and take the limit $\De \to 0.$ As a result one  derives from 
(\ref{nlslq2}) the Lax operator of the sine-Gordon field model
\be{\cal L}= im({u_t } \si ^3 +( k_1 \cos u \si ^1 +k_0 \sin u \si ^2)  
, \ll{Lsg}\ee 
where
the mass parameter $m=m(x,t)$ now 
is not a constant as in the standard case, but
an arbitrary function of $x,t$. The variable mass
also enters the Hamiltonian of this  novel 
  {  sine-Gordon 
 model} as 
\be{\cal H}= \int dx \left [ m(x,t) (u_t)^2 + (1/m(x,t)) (u_x)^2 +
 8(m_0-m(x,t)
\cos (2 \al u )) \right], \ll{msg1}\ee 
which is integrable both at classical and the quantum level for the 
 arbitrary mass
function $m(x,t)$.
 Note that if the mass is independent of time and depends only on the space
coordinate: $m=m(x)$, one can
formally convert the evolution equation  into  the standard sine-Gordon
through a coordinate change:$ x \to X=\int^x m(y) dy  $ and can find its
exact
soliton solution
as
\be u=2\tan^{-1}[ \exp (\gamma \int^x m(y) dy +v t)], \ll{sol} \ee 
which exhibit intriguing structure depending on the choice of the
mass-function $m(x)$. 
 Such 
variable mass sine-Gordon equations may arise  in physical situations
\c{msg} and therefore the related exact results become  important.
\subsection
{Inhomogeneous  NLS model}
Nonlinear Schr\"odinger equation belongs to the nonrelativistic class.
Therefore we should use the rational $R$-matrix and the rational L-operator
(\re{LK}) with suitable realisations of algebra (\re{k-alg}).
A simple such realisation may be given by 
considering site-dependent values for central elements in (\ref{LK}) and in
the
  generalized HPT 
 \be
 s^3=s-N, \ \ \ \    s^+= g_0(N) \psi, \ \ \  s^-= \psi^\dag g_0(N)
, \ \ \ \ g_0^2(N)=m^-+m^+ (2s -N), \ \ N=\psi^\dag \psi.
\ll{ilnls} \ee
This  exactly integrable 
quantum discrete model is an inhomogeneous generalisation of the known
lattice NLS \c{korepinsg}. In the continuum limit one may introduce 
  the inhomogeneity by choosing the eigenvalues of the central elements as
$$ c^1_1={1 \ov \De} +f, c^1_2= -({1 \ov \De} -f),  c^0_1=-c^0_2=g$$
with $f$ and $g$ being  space-time dependent arbitrary functions.
The Lax operator of the field model would  be given formally 
by that of the NLS model, where  
the constant spectral parameter  should be replaced by $\ti \la= g \la + f$
and the field variables by $\psi /\sqrt g.$
 Particular choice of these 
functions as $f={4 x \ov t}, g={1 \ov t} $ would yield integrable  
 { cylindrical NLS}
 \c{rl} like equation at the quantum level.

\subsection
{Inhomogeneous  toda chain}
Another interesting  
 realisation of algebra 
(\re{k-alg}) may be given by
\be
 s^3 =-ip, \ s^\pm=  \al e^{\mp u } \ \ 
 \ll{dtl1} \ee
with $m^\pm=0, $ which leads to  the construction
of  quantum 
Toda  chain model.  A consistent choice  of the Casimir eigenvalues like  
 $c^0_2=c^1_2=0$ together with $ c^0_1$ and $c^1_1  $  taken as space-time dependent
coefficients $c^0_j(t)$ and $c^1_j(t)  ,$ would now result a novel quantum
integrable 
 inhomogeneous Toda chain given by the 
 Hamiltonian
 \be \ H= \sum_j (p_j +{c^1_j \ov
 c^0_j})^2+{1 \ov c^0_jc^0_{j+1}} e^{u_j-u_{j+1}}. 
\ll{todah}\ee
For $ c^1_j=0$  the evolution equation can be written down in an interesting
 compact form
\be 
{\partial^2 \ov \partial t^2}
  u_j=
e^{u^+_j-u^-_{j+1}}-e^{u^+_{j-1}-u^-_{j}},\ll{ihtoda}\ee
where $u^\pm_j(t)=u_j(t)\pm \phi_j (t),$ with $\phi_j(t)$ being an
 arbitrary function inducing inhomogeneity in the system. 
Note that, when the $c'$s are time independent
coefficients such  inhomogeneities  can not be removed through gauge
transformation or variable change.

 Using similar procedure one may construct impurity spin chains
 in a different  way 
by seeking various spin
operator realisations of the  Lax operators (\ref{nlslq2}) or (\ref{LK})
at the impurity sides.


\section {Concluding remarks}

Thus we have    prescribed  a systematic  scheme for 
 constructing  a novel series of inhomogeneous quantum
integrable models belonging to  the
 lattice as well as the field models
of both relativistic  ( $q\neq 1$) and nonrelativistic 
 ($q=1$) class
 along with their corresponding classical counterparts. 
The scheme is based on an algebraic approach, 
where the generators through different realisations construct 
nonlinear functions   of field operators and the Casimir operators 
 with space-time dependent  eigenvalues introduce inhomogeneity 
into the system. 
In our scheme
one also obtains automatically  
the  Lax operators and the $R$-matrices of the 
models constructed.

\ni {\bf Acknowledgment}: 

I thank the
Humboldt Foundation, Germany and the organisers of NEEDS99 for financial
support. 

 \end{document}